\documentclass[a4paper, DIV=15]{scrartcl}

\usepackage{indentfirst}

\usepackage[inline]{enumitem}
\usepackage[nonumberlist]{glossaries}

\usepackage[numbers]{natbib}
\usepackage{compactbib}

\newglossaryentry{ballot sheet}{%
  name={ballot sheet},
  description={A list of candidates standing for election and used by a
    voter to cast their vote; can be in paper or digital form, or mixed}
}

\newglossaryentry{candidate}{%
  name={candidate},
  description={A person standing for election to become a representative
    for their constituency; candidates are eligible for the quorum (not
    the usual meaning of quorum)}
}

\newglossaryentry{constituency}{%
  name={constituency},
  description={\emph{(also known as a \textbf{legislative},
      \textbf{election} or \textbf{voting district,}} \emph{or a
      \textbf{division})} An area in which candidates stand to become
    the representative for that area; a constituency may have several
    polling stations}
}

\newglossaryentry{dispute}{%
  name={dispute},
  description={A process in which a quorum can challenge some votes; if
    the challenge is successful, the votes are excluded from the count}
}

\newglossaryentry{election}{%
  name={election},
  description={\emph{(also known as a \textbf{ballot})} A process in
    which a group of voters cast votes for candidates; the candidates
    with the most votes become the representatives for the constituency.
    For simplicity, the word ``election'' is also use for a referendum}
}

\newglossaryentry{election instance}{%
  name={election instance},
  description={A software representation of the election}
}

\newglossaryentry{identity provider}{%
  name={identity provider},
  description={A software app or a person who generates a unique
    reference for each voter, which is incorporated into the voter's
    identity string}
}

\newglossaryentry{instance}{%
  name={instance},
  description={A software platform that is a representation of an
    election or a polling station and its data. Each such entity may
    have more than one instance. They are all identical to each other}
}

\newglossaryentry{key}{%
  name={key},
  description={A public key or a private key for an entity. Together,
    they comprise the key pair}
}

\newglossaryentry{key pair}{%
  name={key pair},
  description={Asymmetric cryptographic key pair or pairs, which
    includes a public key and a private key}
}

\newglossaryentry{log}{%
  name={log},
  description={\emph{(also known as the \textbf{election log})} A full
    record of the election transactions that is available to at least
    all parties involved in the election. It may be publicly available.
    It is organized as a blockchain of messages}
}

\newglossaryentry{message}{%
  name={message},
  description={A record of a transaction in the log, such as the
    creation of an entity}
}

\newglossaryentry{none of the above}{%
  name={none of the above},
  description={An option in a ballot to reject all candidates or all
    options}
}

\newglossaryentry{observer}{%
  name={observer},
  description={A person or other legal entity who monitors the ballot to
    ensure it is free and fair. Observers are eligible for the quorum}
}

\newglossaryentry{operator}{%
  name={operator},
  description={\emph{(also known as an \textbf{election operator})} A
    person or other legal entity who manages an election instance or a
    polling station instance. They must have no other role in the
    election}
}

\newglossaryentry{option}{%
  name={option},
  description={One of the choices presented to voters in a referendum}
}

\newglossaryentry{owner}{%
  name={owner},
  description={\emph{(also known as the \textbf{election owner})} A
    person or other legal entity who sets up the election. They are
    responsible for creating all the software elements of the election
    (election instances, etc.)}
}

\newglossaryentry{polling station}{%
  name={polling station},
  description={A physical polling station is a place where voters can
    create a new ballot sheet and cast their vote. An online or virtual
    polling station is a piece of software, such as a website or app,
    which voters can use to create ballot sheets and cast votes}
}

\newglossaryentry{polling station instance}{%
  name={polling station instance},
  description={A software representation of a polling station}
}

\newglossaryentry{quorum}{%
  name={quorum},
  description={\emph{(also known as an \textbf{election quorum})} In
    this voting system, a quorum is the minimum number of candidates
    plus the minimum number of observers in each constituency needed to
    agree (or not) that the election has been correctly set up and that
    voting was free and fair. Further, a quorum can choose to exclude
    certain votes}
}

\newglossaryentry{referendum}{%
  name={referendum},
  description={A ballot without candidates where voters choose between
    one of several options (often, only two). The quorum comprises only
    observers}
}

\newglossaryentry{registered voter}{%
  name={registered voter},
  description={A person for whom it has been verified that they are
    allowed to vote in the election and who have been registered to do
    so}
}

\newglossaryentry{teller}{%
  name={teller},
  description={A person or software app that verifies a voter's identity
    and registers their vote}
}

\newglossaryentry{timestamp}{%
  name={timestamp},
  description={The difference, measured in seconds, between when a block
    was created and midnight on 1 January 1970 UTC}
}

\newglossaryentry{voter}{%
  name={voter},
  description={A person who has a right to vote in the election}
}

\newglossaryentry{voter identity string}{%
  name={voter identity string},
  description={A unique string generated for a voter when they register
    to vote, which is used to confirm their identity when they vote}
}

\newglossaryentry{write-in}{%
  name={write-in},
  description={{\emph{(also known as a \textbf{write-in candidate},
        \textbf{write-in referendum option} or \textbf{write-in
          option}})} Voters can add the names of candidates or options
    to their ballot sheet so that they can vote for them}
}

\makeglossaries

\begin{document}

\title{Electt: running auditable and verifiable elections in
  untrusted~environments}

\author{Kirill~A.~Korinsky \\
  kirill@korins.ky}

\date{} 

\maketitle

\begin{abstract}
We present a system for running auditable and verifiable elections in
untrusted environments. Votes are anonymous since the order of
candidates on a ballot sheet is random. Tellers see only the position of
the candidate. Voters can check their vote. An election is auditable
using blockchain log. Threshold-encryption, which is used to implement
the quorum, prevents a deadlock from occurring if a minority of
candidates or observers tries to sabotage the election. Candidates and
observers can indicate that the election was free and fair by exposing
their keys, which are used by the system to decrypt each vote. Ballot
sheets are encrypted by onion-routing, which has a layer with the key of
the election instance, so it's impossible for a quorum to decode the
results before they have announced their decision by exposing their
keys. A register of voters ensures that only verified voters can vote
without compromising their identity. If there any doubts about the
identity of a voter, their vote can be excluded from the election, if a
quorum agrees. This system is designed to scale from one instance to a
distributed system that runs over an unlimited number of instances,
which can be achieved using cloud instances or smartphones belonging to
voters or tellers.
\end{abstract}

\section{Introduction}

This document describes Electt, which is a system for running auditable
and verifiable elections or referenda in untrusted environments. It
supports elections using either printed or digital ballot sheets, and
elections using both types of ballot sheet. An election has one or more
constituencies with different candidates and settings.

Votes are kept secret by randomizing the order of the candidates on each
ballot sheet and by encrypting votes. Voters can also choose their
preferred order of candidates, and they can check the system to see that
their vote has been registered correctly. Further, it is possible to run
an election in which votes must be confirmed by the voter on the system
after the vote has been registered.

This system ensures that the results are verifiable using a
blockchain-like election log. Candidates and observers can indicate
their acceptance that the election was free and fair by publishing their
private keys, which are then used by the system to decrypt each ballot
sheet.

An election is auditable since the whole election log is stored on
multiple servers and devices, which are automatically synchronized so
that the log on each device contains exactly the same transactions. Each
record contains the ID of its parent. The ID for a record is the value
from a cryptographic hash function based on what is in the record. This
log is publicly available.

The system uses a mix of onion routing and threshold encryption without
a trusted dealer to implement a quorum. This prevents a deadlock from
occurring if a minority of candidates or observers tries to sabotage the
election.

The system runs on several identical servers, called election instances.
Each instance has its own set of keys. When encrypting votes using onion
routing, the final layer is encrypted with the public key of the
election instances, which makes it impossible for a quorum of candidates
or observers to decode the results of an election before they have
announced their decision to accept or reject the results. Thus, this
decision is made by them before they know the results.

The system uses a register of voters to ensure that only verified voters
can vote without compromising their identity. If there any doubts about
the identity of a voter, their vote can be excluded from the election,
if a quorum agrees.

This system is designed to scale from one instance to a distributed
system that runs over an unlimited number of instances so that there is
no single point of failure.

This system is also designed to support multiple types of elections in
which voters can cast their votes using:
\begin{enumerate*}
\item
  paper and pencil,
\item
  smartphones or computers via the internet,
\item
  special machines at polling stations, or
\item
  any combination of these.
\end{enumerate*}

This document does not specify the cryptosystem to be used, but does
gives some requirements for compatible cryptosystems. Thus, a single
election may use any combination of acceptable cryptographic techniques.

\section{Related Work}

There has been extensive research on different election methods,
election protocols and related topics.

\subsection{End-to-end encryption}
For ballots, there has been a significant amount of research on
end-to-end encryption between the voter and the authority that is
responsible for counting the votes.

The first election method to use end-to-end encryption may be the one
proposed by Chaum in 1981 \cite{lens.org/110-054-267-420-832}. He
utilized anonymous channels that are now called mix nets or mix networks
\cite{lens.org/112-374-363-401-477}. Mix nets are very common and many
protocols have been proposed \cite{lens.org/098-899-410-804-655,
  lens.org/144-523-120-682-304, lens.org/105-229-973-260-320,
  lens.org/099-697-472-441-352, lens.org/086-244-413-372-840,
  lens.org/049-480-115-765-526, lens.org/123-516-506-300-754,
  lens.org/034-926-000-535-199, lens.org/137-777-559-272-498,
  lens.org/161-355-444-302-490, lens.org/128-354-471-350-809,
  lens.org/100-429-307-241-27X, lens.org/020-891-743-269-148,
  lens.org/092-870-003-476-01X, lens.org/091-517-109-507-597,
  lens.org/095-097-024-836-954, lens.org/072-774-173-524-621,
  lens.org/081-232-386-799-450}. This method could be used by a voting
machine in offline voting or by a device used by someone to vote. For
example, Belenios \cite{lens.org/031-615-693-822-491} is built upon
Helios \cite{lens.org/179-260-337-155-888} and is closely related to
Benaloh’s simple verifiable voting protocol
\cite{lens.org/186-630-734-820-166}, which itself was partially inspired
by the Sako--Kilian mix net \cite{lens.org/034-533-033-458-743}.

Other methods are based on blind signatures
\cite{lens.org/029-969-568-772-09X, lens.org/080-893-101-504-789,
  lens.org/084-773-960-179-055, lens.org/042-886-646-557-473,
  lens.org/066-126-749-537-802, lens.org/049-568-236-921-464,
  lens.org/004-376-916-021-102, lens.org/131-145-642-088-372,
  lens.org/032-917-912-984-280, lens.org/025-057-517-706-79X}. In this
approach, a teller can sign message content (such as a vote) without
being able to see the content. This method has several issues
\cite{lens.org/069-750-108-466-838, lens.org/070-038-114-608-501}, which
could compromise the entire election or compromise votes cast at
particular polling stations. Moreover, for this system to work, a voter
must be able to encrypt their vote, which means that it is not
applicable for elections with physical polling stations where votes are
cast using paper and pencil.

The use of a signing authority could be avoided with ring signatures.
However, the challenge is then to prevent voters from voting twice,
which can be solved with linkable ring signatures. Many protocols have
been proposed \cite{lens.org/125-484-404-898-66X,
  lens.org/027-277-779-279-270, lens.org/004-619-954-345-224,
  lens.org/003-866-081-551-257, lens.org/014-108-918-949-694,
  lens.org/033-198-523-629-654}, in which votes signed by the same voter
are correlated without identifying the voter.

Other schemes are based on homomorphic encryption, which was proposed
more than 30 years ago \cite{lens.org/132-533-271-387-728,
  lens.org/022-359-745-985-641}. The efficiency has since been improved
by many protocols \cite{lens.org/078-085-837-303-106,
  lens.org/000-893-080-410-393, lens.org/014-917-024-151-43X,
  lens.org/074-160-350-545-316, lens.org/055-778-707-657-771,
  lens.org/099-426-341-644-577, lens.org/140-207-404-141-905,
  lens.org/165-594-020-332-622, lens.org/166-215-483-818-534,
  lens.org/113-245-911-395-000, lens.org/016-435-802-310-718,
  lens.org/018-100-139-011-33X, lens.org/167-922-124-177-068,
  lens.org/066-284-225-372-916, lens.org/032-497-719-884-044,
  lens.org/003-829-524-903-352, lens.org/003-829-524-903-352,
  lens.org/037-663-871-004-016, lens.org/148-109-293-958-411}.
Receipt-free methods \cite{lens.org/043-544-999-231-803} are based on
sharing a secret \cite{lens.org/003-605-393-599-967,
  lens.org/176-209-838-407-390, lens.org/197-959-308-447-111,
  lens.org/059-384-693-388-999, lens.org/132-533-271-387-728,
  lens.org/022-359-745-985-641, lens.org/102-413-502-946-386,
  lens.org/121-345-385-574-866}. For example, DEMOS-2
\cite{lens.org/057-933-304-024-478}, CHVote
\cite{lens.org/196-286-870-612-145}, Norwegian vote
\cite{lens.org/060-578-962-864-406}, Neuchâtel's cast-as-intended
\cite{lens.org/130-632-380-261-519}, Selene
\cite{lens.org/006-346-808-095-345}, and k-out-of-l voting
\cite{lens.org/060-722-326-760-28X, lens.org/070-015-656-805-284} are
based on the ElGamal encryption scheme, which operates on a cyclic group
with the assumption that a symmetric external Diffie--Hellman key
exchange on these groups is impossible or at least hard enough
\cite{lens.org/007-282-168-135-199}.

Some protocols exploit features of physical ballots:
\begin{itemize}
\item
  Prêt~à~Voter \cite{lens.org/099-360-094-830-145,
    lens.org/036-328-259-525-785, lens.org/000-752-003-543-118,
    lens.org/018-351-571-363-63X, lens.org/085-491-417-622-087}
  randomizes the order of candidates on each ballot sheet. In November
  2014, a state election in Victoria, Australia, used such a method
  called vVote \cite{lens.org/141-060-077-138-815}. A multi-authority
  implementation \cite{lens.org/025-462-923-554-356} uses two separate
  printers \cite{lens.org/172-470-821-552-317} to print each ballot
  sheet without jeopardizing privacy.
\item
  PunchScan \cite{lens.org/092-142-114-115-414} and Scantegrity~II
  \cite{lens.org/053-452-868-859-275} give voters a confirmation code
  after they vote.
\item
  ThreeBallot \cite{lens.org/141-602-118-169-911} preserves voting
  secrecy while verifying the identity of voters by giving each voter
  three ballot sheets. Voters are verified because they have to choose
  one of the three ballot sheets, but they must still trust the
  authority. The divisible voting scheme
  \cite{lens.org/159-243-065-280-06X} is very similar but prints as many
  ballot sheets as the voter would like.
\item
  SplitBallot \cite{lens.org/143-577-255-448-774} uses different ballot
  sheets to split the trust between two parties, creating a sort of
  conflict of interest.
\end{itemize}

Another way to achieve end-to-end encryption is to use a trusted random
number generator. Bingo voting \cite{lens.org/083-362-541-653-506} uses
a random number list. The number of entries in the list is the number of
candidates times the number of voters. At least one approach
\cite{lens.org/104-696-682-700-169} has been proposed that eliminates
the dependency on the trusted random number generator.

The system proposed inherits and combines ideas from some of these
methods. It uses a type of mix net with onion routing
\cite{lens.org/124-645-321-705-234} based on threshold encryption
without a trusted dealer \cite{lens.org/061-133-166-950-21X} to encrypt
ballot sheets. Each layer of the onion routing is encrypted using the
key of a quorum, which is implemented using a threshold cryptosystem.
The final layer is the quorum of election instances, which prevents a
quorum of candidates or observers (or both) from independently
decrypting the result. A ballot sheet can then be decrypted only with
the agreement of each quorum.

\subsection{Auditable logs}

The first system with an auditable log was VoteBox in 2009
\cite{lens.org/004-984-634-864-389}. Bitcoin was introduced in 2008
\cite{lens.org/053-135-640-121-558}. It uses blockchains, which are
based on the work of Haber et al.\ \cite{lens.org/129-991-584-640-755,
  lens.org/084-495-834-478-057, lens.org/008-094-174-654-321}. There is
no proof that VoteBox was based on blockchain technology or whether the
techniques used were developed independently. Various protocols
\cite{lens.org/003-866-081-551-257, lens.org/004-619-954-345-224,
  lens.org/006-517-212-572-749, lens.org/037-241-553-341-287,
  lens.org/058-277-918-081-550, lens.org/060-997-164-815-104,
  lens.org/086-665-777-870-125, lens.org/097-504-782-054-466,
  lens.org/116-663-428-637-322, lens.org/118-644-012-396-288,
  lens.org/141-445-341-285-986, lens.org/141-600-147-710-181,
  lens.org/152-652-521-055-37X}, including the proposed system, have
used auditable logs to ensure that voting is fair and auditable.

\subsection{Distributing the power}

The idea that the power held by a government during a ballot is a threat
was proposed in 1986 \cite{lens.org/022-359-745-985-641} and legitimized
as a privacy concern. To the best of our knowledge, only a limited
number of researchers have considered this concern
\cite{lens.org/060-085-760-908-495, lens.org/014-917-024-151-43X,
  lens.org/003-648-699-600-182}. All of them have suggested splitting
voters when more than one election is held on the same day.

The e-voting system used in Estonia \cite{lens.org/166-429-914-405-840}
does not protect voter privacy or distribute the power. In elections,
all votes are encrypted using the election system’s key, signed by the
key that is embedded into a voter’s national ID card and then sent to
the election authority to be registered in the system. No cryptographic
technology is used, such as mix nets, to protect voter privacy. This
approach is very similar to a masked ballot
\cite{lens.org/052-525-205-580-044}, its improvement a
doubly-masked-ballot \cite{lens.org/085-569-132-247-372}, and other
approaches that use a voter ID card that is distributed before the
election \cite{lens.org/094-472-064-871-063,
  lens.org/073-051-226-070-828}, confirmation codes
\cite{lens.org/080-185-650-100-966}, national ID card, or a
voter-entered secret \cite{lens.org/083-946-101-982-84X}.

One approach \cite{lens.org/086-336-484-137-109} uses majority judgments
and homomorphic encryption, but it is a new form of voting system. Here,
we are interested in implementations based on existing electoral
systems.

ADDER \cite{lens.org/079-618-958-478-653} relies on a single polynomial
key that is shared between several authorities. This approach may fail
if one of the authorities loses their part of the key. Moreover, this
system does not fully distribute power so cannot prevent collusion.

The proposed system inherits many of these ideas. Since any organization
responsible for an election may be corrupt, the proposed system
transfers power to a number of quorums, each of which has a layer in the
onion routing, which, to the best of our knowledge, has not been done
before. Thus, a minority can protect their voice because the quorums are
based on a majority of candidates and possibly, independent observers,
and election instances. To prevent collusion between candidates, votes
are decrypted only after each quorum has agreed that the election was
free and fair. The final layer in the onion routing is encrypted with
the key of the election instances. Thus, it is impossible for a quorum
of candidates or observers (or both) to decrypt the votes before they
have announced their decision. Moreover, everything is auditable via the
log.

It is assumed that traditional pencil and paper systems are less
trustworthy \cite{lens.org/029-128-918-936-28X} than digital systems
based on cryptography, though the challenge remains of convincing the
various stakeholders of the trustworthiness of such systems \cite[p.
  106]{lens.org/059-682-134-117-442}.

\section{Overview of the system}

\subsection{Threshold encryption and quorums}\label{sec:quorums}

Ballot sheets are encrypted using onion routing in which each layer is
encrypted using a group key. The system uses asymmetric threshold
cryptography without a trusted dealer. Each group private key
(Section~\ref{sec:key_pairs}) has $n$ parts and a threshold of at least
$t$ parts must be available for the message to be decrypted, where $t
\leq n$. Each part of the private key is generated by a different entity
(candidate, observer, or instance), which keeps that part secret. Thus,
a trusted dealer is not required to create the private keys or any part
of the private keys.

The threshold acts as a quorum. For each constituency, the onion routing
has two or three layers:
\begin{enumerate*}
\item
  Candidates, if any
\item
  Observers, if any
\item
  Election instances.
\end{enumerate*}
There must be either candidates or observers, or both.

The threshold $t$ for decryption is no less than $1/2$ the number of
entities in a layer $n$ rounded up to a whole number, plus one if the
number of entities is even: $t = \operatorname{floor}(n/2)+1$.

Thus, for each constituency, there is quorum for a majority of
candidates, a quorum for a majority of observers, and a quorum for a
majority of election instances.

The quorums of candidates and observers have several roles:
\begin{enumerate}
\item
To ensure that the election has been set up correctly
\item
To exclude certain ballot sheets registered at particular polling
stations, by particular tellers, or belonging to particular voters
\item
To ensure that the election is free and fair
\end{enumerate}
For each of these, the candidates and observers can agree, disagree, or
abstain.

\subsection{Cryptographic key pairs}\label{sec:key_pairs}

Each key pair has:
\begin{enumerate}
\item
  A public key that is used to encrypt messages and in verification
  messages.
\item
  A private key that is used to decrypt messages and for signing
  messages.
\end{enumerate}

This system has three types of key pairs:
\begin{enumerate}
\item
  Entity keys that are used to sign messages inserted by an entity into
  the log. These keys have $t = n = 1$.
\item
  Group encryption keys that are used to decrypt a layer of the onion
  routing. These keys have $t \geq 1$ and $n \geq 1$, and each part is
  held by a different entity.
\item
  Group signing keys that are used to sign identity strings. These keys
  also have $t \geq 1$ and $n \geq 1$, and again each part is held by a
  different entity.
\end{enumerate}

The system has the following groups:
\begin{enumerate}
\item
  Each layer of the onion routing (candidates, observers, and election
  instances)
\item
  Polling station instances
\item
  Election instances
\end{enumerate}

The keys may be stored in any format, such as:
\begin{enumerate*}
\item
  in memory or in one or more files on a device,
\item
  in a dedicated device like a physical token,
\item
  encoded and printed in some format like a barcode.
\end{enumerate*}

Whatever methods of storing private keys are implemented:
\begin{enumerate*}
\item
  It must be possible to erase them as toxic waste from memory or from
  the physical token.
\item
  It must be possible to identify the owner of a key used to generate a
  signature.
\item
  It must be possible to identify the owner of each part of a private
  key.
\item
  Further, each signature should be detached from the message, which
  means that the signature should not contain the original message.
\end{enumerate*}

Announcing an entity's key means inserting a message into the log
containing the entity's public key. The ID of this message is the key's
ID. The announcement also contains the key type, that is, what type of
entity the key is for.

A group is a set of candidates, observers, or instances. To announce a
group public key, a message is inserted into the log by each group
member. This message contains a list of member IDs, whether it is an
encryption or a signing key, an ID that identifies the part of the
election instance key used, and the values of $t$ and $n$ for this key.
This messages must be signed individually by the member's key. The
number of members in the group is $n$. The group's public key cannot be
used until all members of the group have separately announced it. This
group key also cannot be used if the values of $t$ or $n$ are invalid,
or if $t$ and $n$ are not the same in all announcements.

If the cryptographic system implemented requires additional messages
between members of the group to generate the key, then these messages
are not stored in the election log.

Revoking an entity's key means inserting a message into the log that
contains the key's ID. It is signed by the key that is being revoked.
This key is then no longer valid and any subsequent messages signed by
it will be ignored. The corresponding private key must be erased after
this message has been created.

To revoke a group key, at least $t$ messages, each from a different
group member, that are signed by their entity's key must be inserted
into the log.

A group's private key may have parts that can be extracted from where
they are stored and parts that cannot be extracted (i.e., because they
are on a physical device). To expose a group's private key, the parts
that can be extracted must be inserted into the log as messages signed
by the relevant entity's key. For the parts that cannot be extracted, a
message is inserted that asserts that the physical device has been
transferred to another entity. This entity must insert a message to
confirm that the part has been received and that it is managing the
device.

\subsection{The log}

The log is managed by software called the log manager. This software
runs on election instances, polling station instances, and the bootstrap
server. The log manager and the log can also be implemented on websites
and elsewhere, such as on a voter's device.

The log is available to every party involved in an election, including
candidates, observers, voters, etc. In a general election, the log is
available to everyone.

This log is a full archive of an election and contains all the
transactions for it. The log can be stored as a legal record of the
election.

The log is a chain of blocks. Each block contains:
\begin{itemize}
\item
  block ID
\item
  parent block ID
\item
  block messages
\end{itemize}

The ID of each block is the value from a cryptographic hash function
based on the parent block ID and the IDs of all the messages in the
block.

A block is a list of messages. Each message has
\begin{itemize}
\item
  message ID
\item
  message type
\item
  digital signature
\item
  message content
\end{itemize}

The ID of each message is the value from a cryptographic hash function
based on the parent block ID of the block that contains the message and
on a signature. A message is signed using the key of the entity that is
inserting the message into the log. The message signature is based on
the message type and message content. A message is valid if it contains
a verified signature and its ID is valid for its parent block.

Valid messages are sent to the log manager. There can be one or more
managers. They each have a buffer of new messages that are to be added
to the log. Periodically, the messages in the buffer are formed into a
block and deleted from the buffer. A block is valid only if all the
messages within it are valid and its ID is valid for its parent block.
When a manager creates a new block, it sends the block to all the
managers it knows about, which are called its neighbors. A block is
inserted into the blockchain only if a majority of neighbors accepts it.
A majority is defined as $1/2$ the number of neighbors rounded up to a
whole number plus one, but at most the number of neighbors. If a block
cannot be added, the messages are returned to the buffer.

For each entity, the log may contain one or more messages that are
related to that entity. The message ID of the first message for the
entity is used as the ID of this entity. Usually the first message
announces the entity's public key, which is used to verify the entity's
signature, but if an entity does not have its own key pair, like when
registering a vote, the ID of the message announcing the entity should
be used as the ID for the entity.

\subsection{Trust tree}

Initially, the log is empty. The first block contains only one message,
which announces the public key that is used to verify the signature for
this message. This is called a self-signed message. This key is
automatically trusted and it is called the bootstrap key.

When a public key is announced using a trusted key, then it becomes a
trusted key itself. For example, if the system trusts key A, then it
will trust each key added using A, and all keys added using these other
keys, etc. This results in a trust tree. Any messages signed by an
untrusted or unknown key are ignored.

Each announced key has a limited ability to insert messages and announce
new keys. This limitation depends on the key type. A bootstrap key can
announce candidates, observers, election instances, and polling station
instances, and it can create constituencies and the election. A polling
station instance key can announce tellers, etc. If a key is announced by
the wrong type of key, it will not be accepted as a trusted key and all
messages signed by it will be ignored.

A trusted key can revoke trust from itself or any key in its trust tree.
This is done by adding a revoke message to the log. This message has a
flag to indicate if only one key or the full tree is being revoked.

If a message revokes only one key, then all keys announced by that key
become orphans. They become the root of their own trust trees. Keys in
such a tree can be revoked only by the orphan key's tree.

\subsection{Bootstrapping}

A new election is bootstrapped by an owner via a temporary process on
the bootstrap server that creates the bootstrap key, which is
self-signed.

When an election has been fully set up and all voters have been
registered, the bootstrap key is revoked, and all subsequent messages
signed by it are ignored. When it is revoked, all the keys that have
been announced by it all become orphans and independent. The bootstrap
stage is then finished, and the bootstrap server and the owner are no
longer involved in the election.

The bootstrap key is revoked before the election is approved and before
$T_a$ (Section~\ref{sec:creating_election}).

\subsection{Creating an election}\label{sec:creating_election}

An election is created by someone referred to as the owner using the
bootstrap key. If the election requires additional parameters, such as
its name, these are included in its announcement message.

An election can have the following parameters:
\begin{itemize}
\item
  timestamp when it is created
\item
  list of election instances
\item
  list of constituencies
\item
  $T_s$ voting start datetime
\item
  $T_e$ voting end deadline
\item
  $T_r$ deadline by which all voters should be registered
\item
  $T_a$ deadline by which a quorum can accept that the election has been
  set up correctly
\item
  $T_d$ deadline by which disputes can be created and acted upon
\item
  $T_c$ deadline by which the quorum can accept that the election is
  free and fair
\end{itemize}

These parameters cannot be changed once the election has been approved.

The minimum requirements for an election are:
\begin{itemize}
\item
  at least one constituency
\item
  at least one election instance
\item
  $T_s$, which can be now or in the past (a date in the past means that
  voting starts as soon as the election is approved)
\item
  voting end deadline
\item
  all deadlines should be defined and should satisfy:
  \begin{equation*}
    T_r < T_a < T_e < T_d < T_c
  \end{equation*}
\end{itemize}

When the election is announced, the election instances create and
announce the group signing and encryption keys
(Section~\ref{sec:quorums}).

An election can be approved only when all the candidates and observers
in each constituency have announced their group encryption key.

\subsection{Creating a constituency}\label{sec:create_a_constituency}

An election has one or more constituencies. A referendum has one
constituency. In an election, candidates stand in constituencies, which
return a number of representatives. In an election for the officers of
an organization, each post (president, treasurer, secretary, etc.) is a
constituency. A voter may be allowed to vote in one or more
constituencies.

Constituencies are created by the owner using the bootstrap key. If
additional parameters are required for a constituency, such as a name,
these are included in the announcement.

A constituency can have the following parameters:
\begin{itemize}
\item
  list of candidates or, for a ballot asking a question, the list of
  options (or answers)
\item
  list of polling stations
\item
  list of observers
\item
  list of mandatory identity providers for voters
\item
  list of parameters for identity providers
\item
  number of candidates in the quorum
\item
  number of observers in the quorum
\item
  number of representatives to be elected
\item
  whether it supports write-in candidates or write-in referendum options
\item
  whether it supports ``none of the above''
\item
  whether confirmation of a vote is mandatory or optional, or whether
  votes are not confirmed at all
\item
  weight of each ballot sheet:
  \begin{itemize}
  \item
    null, which means that the weight is specified individually for each
    voter
  \item
    any positive non-zero real number
  \end{itemize}
\item
  voting system (see below)
\item
  the overall maximum number of votes (see below)
\end{itemize}

A vote means each selection of a candidate or a referendum option on a
ballot sheet. There may be multiple votes for a ballot sheet.

A voting system has the following options for each ballot sheet:
\begin{itemize}
\item
  minimum number of votes (natural non-zero number)
\item
  maximum number of votes (natural number, greater than or equal to the
  minimum)
\item
  whether votes are ordered (as in systems like STV)
\item
  if the maximum number of votes is more than one, whether a voter is
  allowed to vote for the same candidate or option multiple times
\item
  whether each vote has the same weight or whether the weight is
  distributed evenly over the votes
\end{itemize}

The overall maximum number of votes can be:
\begin{itemize}
\item
  Overall for the constituency: When it is reached, any subsequent votes
  are ignored.
\item
  Per polling station: When it is reached, any subsequent votes are
  ignored for that polling station.
\item
  For particular polling stations: When it is reached, any subsequent
  votes are ignored for that polling station (this maximum may be higher
  than the per polling station limit but cannot be higher than the
  overall maximum).
\end{itemize}

If a constituency does not have a specified maximum number of votes,
then the number of votes cast in that constituency cannot be more than
the number of voters registered to vote in it.

The minimum requirements for constituencies are:
\begin{itemize}
\item
   The minimum number of candidates or options is one of:
  \begin{enumerate*}
  \item
    at least two candidates or options, so that an election with one
    candidate or one option is invalid,
  \item
    one or more candidates or options with ``none of the above''
    enabled, or
  \item
    zero or more candidates or options with write-in candidates or
    options enabled
  \end{enumerate*}
\item
  at least one polling station
\item
  at least one mandatory identity provider
\end{itemize}

Observers are mandatory for a referendum with no candidates but are
optional for an election, which may be run without them.

As soon as a constituency is announced and before the election is
approved, all relevant candidates and observers should create and
announce the quorum (Section~\ref{sec:quorums}).

\subsection{Creating candidates and observers}

Candidates and observers are created in a similar way to each other:
\begin{enumerate}
\item
  Create a new key pair for the entity.
\item
  Send a request to announce the entity's public key to the owner.
\item
  Wait until the public key is added to the log.
\item
  Announce the entity with a message signed by its key and include its
  name. If an election requires additional parameters for the entity,
  these are included in this message.
\end{enumerate}

If an entity contains a mistake or its private key has been compromised,
it can be destroyed to prevent future usage. This is done by adding a
revoke message to the log for the entity's key, signed by the owner's
key or the entity itself.

\subsection{Creating instances for the election and polling stations}

The election and each polling station can run on one or more instances.
An instance is a software representation of the election system or
polling station plus all the relevant data. The instances are identical
to each other, which achieves fault tolerance through redundancy.

Management of the instances is fully automated, and under the control of
an impartial operator who is independent of the candidates and
observers.

An instance may be in the cloud or on a local server. The private key of
a local server is stored on a physical token, which is held securely by
the instance operator. The private key of a cloud instance is also held
securely by the instance operator and may be stored on a physical token.

If candidates and observers have a physical token that stores their
private key, they can take it to any election instance to get that token
used, which allows them to expose their key.

An instance is created as follows:
\begin{enumerate}
\item
  Create a new key pair for the instance.
\item
  Send a request to announce the entity's public key to the owner.
\item
  Wait until the key is added to the log.
\item
  Announce the instance with a message signed by its key and include its
  name and its type. If an election requires additional parameters for
  the instance, these are included in this message.
\end{enumerate}

If an instance contains a mistake or its private key has been
compromised, it can be revoked to prevent future usage. This is done by
adding a revoke message to the log for the instance's key signed by the
owner or the instance itself.

\subsection{Creating polling stations}\label{sec:createPS}

A polling station does not have its own key pair. Instead, it uses any
of the keys of its polling station instances.

Polling stations are created as follows:
\begin{enumerate}
\item
  The polling station is announced by the election owner in a message
  signed by the bootstrap key. The announcement includes the name of the
  polling station and a list of polling station instances. If an
  election requires additional parameters for the polling station, these
  are included in this message.
\item
  Wait until the announcement is added to the log.
\end{enumerate}

If the data for a polling station contains a mistake, the polling
station can be destroyed to prevent further usage. This is done by
adding a revoke message to the log for the polling station, signed by
the owner.

A polling station instance can belong to only one polling station. If a
polling station is destroyed, its instances can be reused by a new
polling station.

As soon as a polling station is announced, its instances then create a
group signing key and they separately announce it
(Section~\ref{sec:quorums}).

\subsection{Identity providers}

The system supports different types of identity provider:
\begin{itemize}
\item
  An official, such as a civil servant of the government running the
  election or an officer of an organization running an election
\item
  All the tellers collectively comprise a single identity provider
\item
  Software that is not part of this system, for example, an official
  authorization service
\end{itemize}

Different elections use different methods to verify someone's identity.
When a person registers to vote, their identity can be checked by one or
more identity providers. Each identity provider generates a unique
reference for each voter as follows:
\begin{itemize}
\item
  If the person is registering for a postal vote, then their name and
  address can be used as the unique reference.
\item
  If the person is presenting their passport, ID card or other ID in
  person, then the number from the ID is the unique reference.
\item
  Software can be used to generate a unique reference, such as through
  the SSO or SAML protocols.
\end{itemize}

The unique reference for a person for an ID provider is fixed,
regardless of which software instance or which employee within an
identity provider generated the unique reference.

A voter may be identified by any identity provider but must be
identified by all mandatory identity providers. They can get their
identity verified in different ways, and for each method, an identity
string is generated for each relevant election or polling station
instance. An identity string is generated as follows:
\begin{enumerate}
\item
  The ID of the constituency, the ID of the identity provider and the
  unique reference generated by the identity provider are concatenated
  into a single string.
\item
  This string is signed by the group signing key of a relevant entity,
  such as a polling station or election instance.
\item
  The final identity string is this digital signature.
\end{enumerate}

Each identity provider has:
\begin{itemize}
\item
  ID, which is the ID of the message announcing the identity provider
\item
  list of polling stations, which may be empty
\item
  other data, such as the name of the identity provider, as required by
  the law
\end{itemize}

If the list of polling stations is not empty, then the identity provider
can verify identities only for people voting at those polling stations.
Otherwise, it is a global provider and can be used only by election
instances.

An identity provider is created as follows:
\begin{enumerate}
\item
  Announce the identity provider using the bootstrap key.
\item
  Wait until the announcement is added to the log.
\end{enumerate}

If an identity provider contains a mistake it can be revoked to prevent
future usage. This is done by adding a revoke message to the log for the
identity provider's ID signed by the bootstrap key.

An identity provider can be registered or revoked only after the
election has been created and before $T_r$.

If an identity provider is revoked, all voters with identity strings
from this identity provider are revoked.

\subsection{Voters}

The system maintains a register of voters, i.e., those people who are
allowed to vote. The register also lists the polling stations at which a
voter may cast their vote.

Each voter has:
\begin{itemize}
\item
  ID, which is the ID of the message announcing the voter
\item
  vote weight (a brief description of how this is used appears in
  Section~\ref{sec:create_a_constituency}); this weight is ignored if
  the constituency has a non-null weight
\item
  constituency ID
\item
  a list of identity strings
\end{itemize}

A voter is created as follows:
\begin{enumerate}
\item
  Create identity strings for each identity provider for the voter's
  constituency.
\item
  Announce the voter using the bootstrap key.
\item
  Wait until the announcement is added to the log.
\end{enumerate}

Voters can be registered or revoked only after the election has been
created and before $T_r$.

If a voter identity string contains a mistake, it can be revoked to
prevent future usage. This is done by adding a revoke message to the log
for the voter's ID signed by the bootstrap key.

\subsection{Approving the election}

Once the election has been created and all voters are registered, the
election next needs to be approved, which confirms that it has been set
up correctly. The candidates and observers in each constituency do this
by individually:
\begin{itemize}
\item
  Creating a message saying that they agree that the election has been
  set up correctly.
\item
  Creating a message saying that they disagree that the election has
  been set up correctly.
\item
  Abstaining by not creating a message.
\end{itemize}
Each of these messages contains a timestamp and signature created by the
group signing key for the constituency.

If a candidate or an observer is a member of more than one constituency,
they can vote in each of them to approve the election or not.

Only the first message inserted by a candidate or observer in each
constituency, after the owner has revoked their key, is counted. All
subsequent messages are ignored, as are all messages inserted before the
owner revoked their key.

If the quorums in all constituencies agree that the election has been
created correctly and if the members of all constituencies have
announced their group keys, then the election will proceed. This can be
announced by any election instance with a message that contains a
timestamp. If the log has more than one such message, only the first is
accepted the subsequent messages are ignored.

If the quorum in any constituency disagrees that the election has been
set up correctly or if a quorum is not reached by the deadline for any
constituency, then the election will not proceed. All election instances
will revoke their keys, and all subsequent messages will be ignored.

If the log contains an election-created message but at least one key has
been revoked by an election instance, the log has been compromised and
this election cannot proceed.

If the bootstrap key has not been revoked by $T_a$ for the election,
then the election will not proceed.

Once the election has been approved in this way, the bootstrap stage is
finished. The election parameters can no longer be changed. In addition,
candidates, observers, polling stations, constituencies, identity
providers and voters can no longer be added or revoked.

\subsection{Tellers}

Each physical polling station requires at least one teller, who
registers the votes cast in person. For a virtual polling station, the
tellers are software instances. For postal votes, the teller may be an
official of the organization running the election or may be software,
etc.

A teller is created for a polling station as follows:
\begin{enumerate}
\item
  Create a new key pair for the teller.
\item
  One of the polling station's instances announces the teller's key.
\item
  Wait until the public key is added to the log.
\item
  Announce the teller with a message signed by its key and include its
  name. If an election requires additional parameters for tellers, these
  are included in this message.
\end{enumerate}

If a teller contains a mistake or its private key has been compromised,
it can be revoked to prevent future usage. This is done by adding a
revoke message to the log for the teller's key signed by the polling
station instance's key or the teller's key. This does not affect the
signature of any voter ID or any votes already registered by the teller.
However, the teller can no longer register new votes, and all voter IDs
that have been signed by the teller's key but who have not yet
registered a vote will be ignored.

Tellers can be created or destroyed from when the election is approved
until voting is finished.

\subsection{Ballot sheets}

In this system, each voter can choose to have a paper ballot sheet, an
electronic ballot sheet or a mixed ballot sheet. A paper ballot sheet
has two parts:
\begin{itemize}
\item
  A private part that lists the candidates in some order, the unique
  ballot sheet ID, and the ballot sheet's private key.
\item
  A public part on which the vote is made. It does not have the
  candidates' names or anything that may be used to identify the
  candidates, and it does not indicate whether the vote has to be
  confirmed. It contains only the ballot sheet ID, which may be encoded
  as a barcode.
\end{itemize}

The paper version is easy to tear apart so that the private part can be
kept secret.

A mixed ballot sheet has the list of candidates in electronic form, but
the public part is printed out.

A ballot sheet can be created by facilities at a polling station
(computer or polling machine and a printer) or online. A voter can use
any device with an internet connection, such as a smartphone or
computer. If the voter uses their own device, the private part of the
ballot sheet is generated by that device, and stored only on this
device.

Each voter ID can be used in only one constituency. Since ballot sheets
are for only one voter ID, then if a voter can vote in more than one
constituency, a separate ballot sheet (and voter ID) is created for
each.

The payload for a ballot sheet contains:
\begin{itemize}
\item
  whether this vote should be confirmed
\item
  list of candidates or options, as they appear on the ballot sheet
\item
  signature for both fields using the ballot sheet's key
\end{itemize}

The payload of each ballot sheet is encrypted. The payload is used to
map the vote cast (which is just the position of a candidate or option
on the ballot sheet) with the list of candidates (or options).

A ballot sheet is created as follows:
\begin{enumerate}
\item
  A key pair is generated for the ballot sheet, which is used only to
  confirm or revoke a vote.
\item
  Announce the ballot sheet's key with a polling station instance's key.
\item
  The voter can choose to order the candidates randomly or chose an
  order. They can choose to omit some candidates so that the list is
  shorter. They can also add the names of any write-in candidates. They
  can also add null names to pad out the list of candidates. The number
  of candidates on a ballot sheet must be at least the number of
  representatives returned.
\item
  The ballot sheet is announced with a message containing its encrypted
  payload and the constituency ID.
\item
  A ballot sheet can be printed out if required.
\end{enumerate}

For a constituency, a voter can generate as many ballot sheets as they
like at different polling stations, but they can use only one, which can
be any of them.

\subsection{Encrypting ballot sheets}

This system uses both onion routing and threshold encryption to encrypt
a ballot sheet's payload.

In onion routing, a message is encapsulated in several layers (or hops),
analogous to the layers of an onion. Each layer is encrypted with a
different public key, so the message can be fully decrypted only by
using all the private keys in the right order. If any one key is
unavailable or they are used in the wrong order, then the message cannot
be decrypted. This ordered list of keys is called a path.

Each path contains the keys for all layers for a constituency:
candidates, observers, and the election instance. This path is generally
unique for each constituency and has at least two hops:
\begin{enumerate*}
\item
  the election instance
\item
  and candidates or observers.
\end{enumerate*}
If a constituency contains candidates and observers, the path has three
hops.

When a ballot sheet is created, the payload is encrypted on the device
that generated it.

\subsection{Casting a vote}

When a person votes, their identity is first checked by a physical or
virtual teller from any polling station for this constituency whose key
has not been revoked. The teller will generate an identity string for
each mandatory identity provider for this constituency. If there are no
mandatory identity providers, the teller will generate an identity
string for an optional identity provider using the ID provided by the
voter. If all these generated identity strings match those on the
register for the voter, then their identity has been confirmed and they
are given a voter ID signed by the teller's key. This voter ID might be
in physical form (printed out) or in digital form.

To ensure their privacy and to exclude any possibility that a teller can
track their real identity using the voter ID, a voter can use two
different tellers to register their vote:
\begin{enumerate*}
\item
  the first to obtain a voter ID and
\item
  the second to register their vote.
\end{enumerate*}

A vote is made using the public part of a ballot sheet. The vote is
marked on it, either physically or electronically, depending on the
voting system used (an X, with 1, 2, 3, \ldots, etc.). Votes can be cast
in person or by post, email, SMS, phone, or online.

The teller will then register the vote by creating a message based on
information from only the public part of the ballot sheet and using the
voter ID. The message is signed by the teller and contains:
\begin{itemize}
\item
  ballot sheet ID
\item
  positions of selected candidates, with a number indicating the
  preference order if relevant or distributed weight
\item
  voter ID and teller's signature
\item
  signature of teller who obtained this ID
\end{itemize}

If a vote is registered for the same ballot sheet or for the same voter
more than once, the system will use only the first. If a vote is
registered by a teller in a polling station that is not in the voter's
constituency, the vote will be ignored.

A voter can check their vote online using the ballot sheet ID or their
voter ID. Until the ballot sheet is decrypted, the information shown is
only the voter ID and the positions in that ballot sheet of the
candidates who have been voted for, with a number indicating the
preference order, if relevant.

When a ballot sheet is created, the voter may specify that they would
like to confirm their vote, or this may be mandatory for a constituency.
In this case, their vote will be counted only if the log contains a
confirmation from the voter and it has not been revoked. Only the first
confirmation or revoke message will be used. If a voter confirms or
revokes a vote for a ballot sheet that does not need to be confirmed,
this message will be ignored.

If the teller registers the vote incorrectly, the teller can revoke it.
However, the teller will be unable to register the vote again for either
this ballot sheet or the voter.

\subsection{End of voting}

Voting ends at $T_e$ or when all voters in all constituencies have
voted, whichever occurs first.

The parts of each group signing key are held by election and polling
station instances until voting is finished, after which they are erased
as toxic waste. If an instance cannot erase its part of a group signing
key, then it should treat the part as if it has been compromised. If the
number of parts compromised for a group is ${\geq}t$, then all votes
cast by voters with an identity string signed by the group should be
excluded from the election via the dispute mechanism. These ballot
sheets will not be decrypted.

All election instances will insert a voting-finished message with a
timestamp into the log to prevent further votes from being registered.

However, at the deadline, some polling station instances may still have
votes in their buffer. These votes will then be inserted into the log.
Such votes will be counted provided that they were registered (i.e.,
inserted into the buffer) before the deadline. The polling station
instance will then confirm that the election has finished by inserting a
voting-finished message with a timestamp to prevent further voting.

The voting-finished messages have flags that confirm that the instance
has erased its part of the group signing key, or not.

\subsection{Disputes: Excluding a polling station, teller, or voter}

In some situations, all the votes registered by a particular teller, at
a particular polling station, or by a voter may need to be excluded.
This process can be started only after at least one voting-finished
message has been inserted into the log and before $T_d$.

In that period, any candidate or observer can insert a message into the
log for their constituency containing:
\begin{itemize}
\item
  a timestamp
\item
  free text explaining why the exclusion is necessary
\item
  a list of polling stations that should be excluded, a list of tellers
  that should be excluded, or a list of voters that should be excluded,
  or any combination of these
\item
  whether the exclusion is due to a leak of private data from a polling
  station instance and the number of suspected instances
\end{itemize}

After this message has been inserted into the log, the system will wait
for the decision of the candidates and observers. The decision is to
agree, disagree, or abstain, which is inserted as a message with a
timestamp into the log.

If any observer or candidate exposes or revokes their key before they
have given their decision on whether to exclude these votes, then the
system will consider that:
\begin{itemize}
\item
  they agree that the votes should be excluded if they have exposed
  their key
\item
  they disagree that the votes should be excluded if they have revoked
  their key
\end{itemize}

The relevant votes will be excluded only if a quorum agrees within $T_d$
that they should be excluded.

If the votes from a polling station are excluded, all further messages
from its instances are ignored. If the votes from a polling station are
excluded due to a leak of private data and if the number of suspected
instances ${\geq} t$, then the votes cast for all voters that contain
identity strings for that polling station are also excluded.

This stage is not finished until each non-excluded polling station
instance has confirmed that it has inserted all votes in its buffer into
the the log and that it has destroyed its part of the group signing key.
If this message cannot be added by a polling station, then to bypass
this daedlock, the quorum should exclude it as if there were a leak of
private data. Otherwise this would be considered as a disagreement by
the quorums in all related constituencies when deciding whether the
election was free and fair.

\subsection{Free and fair}

Once voting has stopped and all disputes about excluding votes are
settled, the candidates and observers must confirm that the voting was
free and fair. Each member of the quorum in each constituency can:
\begin{itemize}
\item
  Agree that voting was free and fair by exposing their part of the
  group encryption key.
\item
  Disagree that voting was free and fair by revoking their part of the
  group encryption key.
\item
  Abstain by not sending a message.
\end{itemize}
All of these messages have a timestamp.

If a key is not exposed before the deadline for this decision, then the
system will consider that this candidate or observer has abstained.

For each candidate or observer, only the first message in the log is
counted. All other such messages are ignored.

The entire election is not considered to be free and fair if any of the
following conditions apply:
\begin{enumerate}
\item
  If ${\geq} t$ election instances have not confirmed that they have
  destroyed their parts of the group signing key by the voting end
  deadline.
\item
  If for a polling station:
  \begin{enumerate*}
  \item
    it has not been excluded by the dispute mechanism,
  \item
    its instances have not confirmed that it has inserted all the
    registered votes from its buffer into the log, and
   \item ${\geq} t$ instances have not confirmed that its part of the
     group signing key has been erased by $T_d$.
  \end{enumerate*}
\item
  If the quorum in any constituency disagrees that the election was free
  and fair.
\item
  If a quorum is not reached by the free-and-fair deadline for any
  constituency.
\end{enumerate}

If the election was not free and fair, then each election instance will
destroy its part of the group encryption key. It will insert a message
indicating that the election has been rejected and confirm that it has
destroyed its part of the group encryption key. Because of the
encryption method, it is then impossible for the ballot sheet payloads
to be decrypted, so that the result can no longer be calculated.

\subsection{Calculating the results}

However, if the election was free and fair, then the result will be
calculated as follows. All election instances that have access to a
quorum of exposed keys will insert a message indicating that the
election was free and fair and will then decrypt the ballot sheets. Once
a ballot sheet has been decrypted by an election instance, it will not
be decrypted again by another instance. In this way, extracting the
ballot sheets can be shared among election instances.

As soon as all non-excluded ballot sheets have been decrypted, all
election instances should erase their part of the group encryption key
as toxic waste.

If the log contains at least one message from an election instance
asserting that the election was free and fair, all rejected messages
should be ignored.

A valid vote
\begin{itemize}
\item
  has a valid signature
\item
  has a valid voter ID and a valid signature for the voter ID
\item
  is the first vote for a voter
\item
  has not been revoked and has been confirmed, if necessary
\item
  has not been excluded
\item
  was registered between $\max(T_s,T_a)$ and $T_e$.
\end{itemize}

For each non-excluded vote, the system will insert a message containing:
\begin{itemize}
\item
  ballot sheet ID
\item
  decrypted ballot sheet payload
\end{itemize}

An election has a maximum number of votes. Any votes registered after
the maximum has been reached are ignored and the related ballot sheets
will not be decrypted. Similarly, a polling station has a maximum number
of votes and related ballot sheets beyond that maximum are ignored.

Using only valid votes that are not being ignored, all election
instances will calculate the number of votes for each candidate or
option in each constituency based on the voting system, the vote
weights, etc., and taking into account write-in candidates or options.

When each election instance has calculated the results, it will insert
them as a message into the log and then destroy its key as toxic waste.

\section{Threat modeling}

\subsection{Election commission}

This system eliminates the need for an organization that controls all
aspects of an election, like an election commission, by splitting its
power into different independent roles:
\begin{itemize}
\item
  The owner, who creates the election and does all the necessary
  administration.
\item
  A quorum of members, who ensure that the election has been set up
  correctly and that voting was free and fair. A quorum can exclude some
  votes via the dispute mechanism.
\item
  Election instances, which are automated systems. They prevent a quorum
  from making decisions about the election behind closed doors. This
  software is managed by operators only.
\item
  Operators, who are not involved in the election as candidates,
  observers, or owners. They are technicians who manage an election
  instance during the election.
\item
  Tellers, who verify the identity of voters.
\end{itemize}

If a political party or other group wishes to manipulate the election,
they need access to more than half of the private keys of the candidates
and observers. This may include the keys of their opponents. They also
need a majority of the private keys of election instances, but these
keys are held securely and only operators have access to them. An
election instance key may be contained within a physical token, so it
would have to be removed from the instance, though instances are
publicly monitored. Moreover, the software on the election instance will
generate an alert to indicate that the key is not present. However, it
is possible that a majority of candidates and observers are under the
control of some organization, such as a political party. This type of
threat cannot be eliminated.

This system uses threshold encryption without a trusted dealer. Thus,
private keys are not generated by or shared with a trusted dealer. This
eliminates the risk that a secret required for decryption is copied
before a quorum reaches a decision on whether the election was free and
fair.

The software installed on instances and the firmware for physical tokens
will be available as open-source software for a reasonable amount of
time before an election. This allows independent researchers to check
that the software does not have a backdoor or predefined keys.

\subsection{Protecting voter identities}

Each voter's identity is protected because the system uses a group
signing key to generate an identity string for the voter's unique
reference from each identity provider. Moreover, each identity string is
a signature that is generated by a key that is identifiable. All
identity strings are generated before the election is approved by the
quorum.

A brute-force attack to determine the identities of all voters
associated with an instance would need to steal the keys from a majority
of instances in the group that was used to create the identity strings.
Some of these keys are stored only inside a token. All polling station
and election instances will erase the private keys that they use for
signing from the token or their memory as soon as voting is complete.

Even if an attack is successful, the attackers will get only the unique
references provided by the identity providers, which safeguards the
privacy of voters.

If there are any doubts or suspicions about whether private data have
been leaked from any instance, the quorum can use the dispute mechanism
to exclude from the election the relevant registered votes. This
protects the real votes of people associated with particular voter IDs
from a brute-force attack.

Thus, all polling station and election instances will erase as toxic
waste the secrets that were used to create identity strings. This makes
it impossible to calculate the results of the election until all votes
have been excluded for voter IDs that contain identity strings from
instances that have not erased their secrets.

A voter ID can be used in only a single constituency. Thus, it is
impossible to identify a voter by comparing how they voted in different
constituencies in a multiple-constituency election.

Moreover, voters can choose to get a voter ID from one teller and then
register their vote with either a different teller or a polling station
that supports online voting. Thus, the tellers do not know how they
voted.

\subsection{Coercion and buying votes}

As with any remote method of casting votes, the system is unable to
prevent someone from stealing someone else's vote.

A voter can print or generate as many ballot sheets as they want, but
only the first vote registered in the system will be counted. This
allows someone to appear to comply with coercive demands at no risk by
voting twice, first as their intended vote and secondly to comply with
the coercion. Similarly, a voter can cast a vote using a friend's voter
ID to prevent their real voter IDs from being exposed.

Because voting records are encrypted, it is very difficult to determine
how a voter voted. Once a ballot sheet has been decrypted, however,
someone can compare the public part with the private part to determine
which candidate was voted for. However, it is still not possible to
match the real person with their voter ID.

\subsection{Deniable votes}

All voting systems should ensure that votes are deniable, which this
system achieves by breaking the link between a person and a voter. Each
identity provider returns a unique reference for each person and the
system uses the instances' group private key to generate an identity
string based on the unique reference. All instances erase their group
private keys as soon as voting is complete, which breaks the link
between the person and the voter via the identity providers. Thus, all
votes are deniable because they are not directly linked to the person
who voted.

This system does not use timestamps or time-related fields when
recording votes to prevent an attack based on assumptions about who
voted by spying on a polling station.

Messages recording a vote are sent to the log in batches by the polling
station instances. So, there is an unpredictable delay before a message
is sent to the log. Thus, it is hard to identify which voter cast which
vote based on when the vote was registered.

\subsection{Identity provider fraud}

Voters can be verified using more than one identity provider. This
prevents collisions and protects against a fraudulent identity provider.
However, the system is vulnerable if all identity providers return
incorrect verification results for a person due to fraud, censorship, or
similar.

\subsection{Cracking votes}

Each ballot sheet contains an unspecified number of candidates or
options in an unspecified order, neither of which can be predicted.

For example, the order of candidates can be randomized using a random
number generator on a voter's trusted device or by a device at a polling
station. The order could be selected by the voter. A voter can remove
one or more candidates from a ballot sheet (provided enough remain for
the voter to be able to cast all their votes). Finally, a voter can add
null candidates. These cannot be voted for, but affect the numbering of
candidates on the ballot sheet. All of these measures increase the
entropy of the payload.

The entropy is further increased because a ballot sheet does not contain
any information about where it was generated or how many candidates it
contains. This makes an attack based on predicting random numbers much
more difficult (if even feasible at all).

\subsection{Excluding inconvenient polling stations}

In many elections, there are exit polls, which can be used to predict
the result of the election.

If a candidate expects that the results from a particular polling
station will not be favorable to them, they cannot exclude any votes
from that polling station unless they have the support of a quorum,
which comprises a majority of candidates and observers for the
constituency.

They can challenge the votes cast at a polling station if they suspect
fraud, but excluding these votes requires a quorum. Such challenges are
out of scope for this system.

\subsection{Predicting the result of an election}

It is impossible to predict the result by analyzing the log, since only
the position of a candidate on a ballot sheet is recorded, the list of
candidates is different for each ballot sheet, and the list of
candidates is encrypted.

Each ballot sheet has its own key pair, but this key pair is used only
to confirm or revoke votes cast for this ballot sheet. It cannot be used
to decrypt a ballot sheet's payload.

If the private part of a ballot sheet is intercepted (e.g., by finding a
paper version in a bin), then it is possible to compare the vote in the
log with this private part. However, the eavesdropper does not know
whether the vote has to be confirmed, and if so, whether it has been
confirmed or revoked. The eavesdropper also does not if this was a
second vote by the voter, which would be ignored.

\subsection{Tampering with the log}

The log is a blockchain. Each block in the chain and the messages in a
block contain the ID of the parent block. Changing a block changes all
subsequent message IDs, since these are based on a hash of the previous
block. Moreover, message ID are used as entity IDs, and changing the IDs
for all entities would change all the signatures in the subsequent
blocks. Thus, changing even a single message in the blockchain would
mean regenerating all subsequent signatures, which would require access
to all the private keys. That seems almost impossible to achieve.

Finally, the blockchain is distributed. Even if the log is changed, some
voters will know their voter ID and some will also have a copy of the
log. Such a voter can easily prove that someone tampered with the log.

\section{Conclusion}

This system eliminates the need for an organization that is responsible
for all aspects of an election, like an election commission. Such an
organization would need to be trusted by everyone. The power, instead,
is transferred to the quorum and to the election software through the
election log. The software is controlled by operators who are
independent of the candidates and observers and have no other role in
the election. The operators have no real power and cannot influence the
results of the election. The decision of the quorum is based on
information that is available to everybody.

This system allows each voter to control how their vote will be
accounted for. Votes can be excluded only if a quorum agrees.

This system does not require any changes to the voting systems we
currently use (first past the post, STV, etc.) and it supports all types
of election, from small and simple votes by corporate stockholders to
referenda to large and complex elections with many constituencies, such
as UK and US general elections. The voting rules can be customized for
each constituency.

Candidates must accept that the election was free and fair before the
result can be calculated, so that it is unreasonable for them to claim
it was not free or not fair after the results have been announced.

Expensive voting machines do not need to be installed at each polling
station since ballot sheets can be created using a cloud polling
station, a smartphone belonging to the polling station, or a voter's
personal smartphone.

This system realizes all of the expected requirements of a voting
platform, such as deniable votes and voter anonymity. Potentially, an
election can be run more cheaply than currently, and it be more
transparent and have a higher degree of trust.

\section*{Acknowledgement}

We thank Jonathan Webley and Kirill Pimenov for comments that greatly
improved the manuscript.

\glsaddall
\printglossaries

\clearpage
\bibliographystyle{unsrtnat}
\bibliography{electt}

\end{document}